# Automated Energy Billing with Blockchain and the Prophet Forecasting Model: A Holistic Approach

Ajesh Thangaraj Nadar
*Department of Computer Engineering*
*SIES GST*
Nerul, India
ajeshraj402@gmail.com

Gabriel Nixon Raj
*Department of Computer Engineering*
*SIES GST*
Nerul, India
gabriel.nixonraj@gmail.com

Nihar Mahesh Pasi
*Department of Computer Engineering*
*PHCET*
Rasayani, India
niharpasi123@gmail.com

Soham Chandane
*Department of Computer Engineering*
*SIES GST*
Nerul, India
szsohamchandane@gmail.com

Yash Arvind Patil
*Department of Computer Engineering*
*PHCET*
Rasayani, India
yasharpmail@gmail.com

Given Name Surname
*Department of Computer Engineering*
*name of organization*
City, Country
email address or ORCID

*Abstract*—The world's reliance on energy imports has turned into a major problem for nations, throwing a great deal of strain on their economy. Researchers and policymakers are under more pressure than ever to develop efficient methods for managing domestic energy resources and fostering economic growth. This study suggests a comprehensive strategy for dealing with problems related to energy supply, demand, and price, with a focus on controlling the spike in peak energy demand. To improve peak load management, a unique technique of feedback and goal-setting is presented. A high-tech system consisting of smart energy meters outfitted with Wi-Fi modules for automation and real-time connection with utility providers is suggested to implement and monitor this mechanism successfully. The system has a user-friendly interface that enables users to access data analysis for more insightful data on energy use. A cost-effective solution for a range of user groups is provided by this innovative design, which also has the ability to promote economic growth and achieve sustainable energy consumption.

## I. INTRODUCTION

Smart Energy Monitoring (SEM) systems have become a game-changing force in the field of energy management thanks to the power of Internet of Things (IoT) technology. These state-of-the-art devices offer real-time information on current and voltage levels, allowing customers to obtain significant understandings of their patterns of power usage. SEM enables people to actively participate in energy conservation and make educated consumption decisions through an easy-to-use mobile application. Remote control of home appliances encourages energy efficiency and optimises usage during times of high demand, creating a more sustainable energy footprint. One unique aspect of SEM is its capacity to predict real-time power bills, providing customers with real-time billing insights that promote awareness and promote sensible usage habits. SEM opens the way to a smarter energy environment by providing in-depth data analysis capabilities, allowing customers to see trends and chances for more efficiency gains. Adopting SEM technology not only simplifies energy management procedures by lowering the number of manual readings, but it also significantly contributes to a future that is greener and more environmentally aware.

## II. STATISTICS

### A. Cost Structure of Distribution Utilities

TABLE 1    COST STRUCTURE OF DISTRIBUTION UTILITIES IN 2019-20(IN RS/KWH)

SOURCE    REPORT ON PERFOMANCE OF POWER UTILITIES 2020-21, POWER FINANCE CORPORATION PRS

| Cost Head | State Sector | Private Sector |
|---|---|---|
| Cost of power | 4.70 | 5.17 |
| Employee cost | 0.51 | 0.49 |
| Interest cost | 0.41 | 0.57 |
| Depreciation | 0.21 | 0.30 |
| Other costs | 0.26 | 0.47 |
| **Total** | **6.09** | **6.99** |

In the State Sector, the cost of power is 4.70 units, employee cost is 0.51 units, interest cost is 0.41 units, depreciation is 0.21 units, and other costs amount to 0.26 units, resulting in a total of 6.09 units. In contrast, the Private Sector incurs higher costs in most categories: cost of power is 5.17 units, employee cost is 0.49 units, interest cost is 0.57 units, depreciation is 0.30 units, and other costs are 0.47 units, with a total of 6.99 units. This data highlights the cost variations between the two sectors in power utilities' financial performance.

GRAPH 1    AVERAGE COST OF STATE ELECTRICTY SUPPLY IN INDIA FORM 2015-20

SOURCE    STATISTA.COM

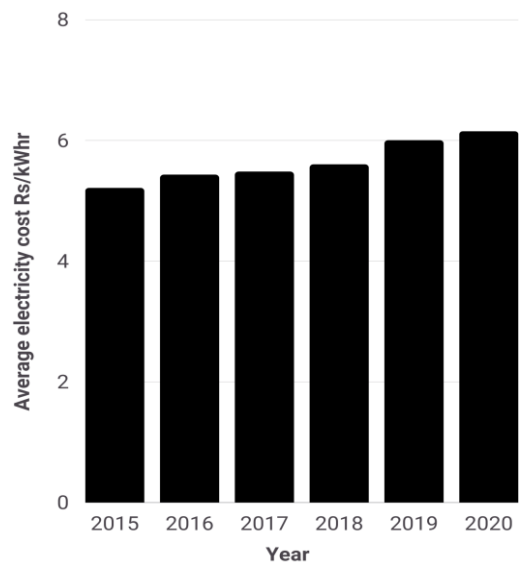

## III. METHODOLOGY & ALGORITHM

In order to create an Internet of Things (IoT)-based electricity energy monitor utilising the ESP32 and integrate it with the Blynk Application, we follow a prototyping approach. This approach's main goal is to develop an automated system that makes it unnecessary to manually read metres and improves real-time power usage monitoring. The SCT-013 Non-Invasive AC Current Sensor and the ZMPT101B AC Voltage Sensor, which precisely measure current and voltage for correct power calculations in kilowatt-hours (kWh), are the system's main parts. While the ZMPT101B sensor assures reliable AC voltage measurement utilising a voltage transformer, the SCT-013 sensor provides non-intrusive AC current measurement up to 100 amperes.

The Prototyping Model is used to quickly construct an operational but unfinished model of the information system. By quickly and crudely building a clone or mock-up of the desired system, this strategy lowers risk. The Prototyping Model is primarily used to demonstrate technical viability, especially when the technical risk is considerable. It also makes it easier to comprehend and elicit user requirements. With the use of this model, we want to reduce risks, keep expenditures in check, and improve understanding of suggested solutions before allocating more funds.

We build and evaluate the IoT-based Electricity Energy Monitor through iterative design and testing phases by using the prototyping approach. To evaluate the system's functionality, efficacy, and viability in real-world circumstances, rigorous experimental testing are carried out under a variety of stress conditions. The information gathered from these tests is used to improve and calibrate the system, assuring its dependability and effectiveness in managing power use. This suggested methodology offers a practical and affordable alternative for promoting sustainability and energy efficiency, enabling users to remotely monitor and manage their energy use.

### A. Architecture

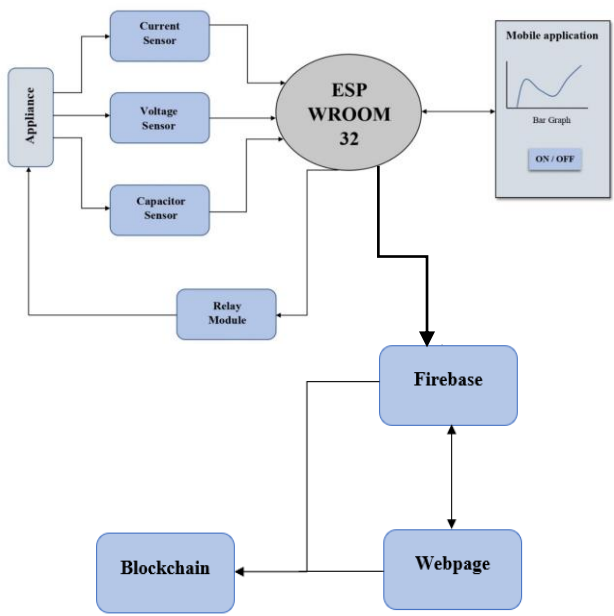

Fig. 1. SEM Flowchart

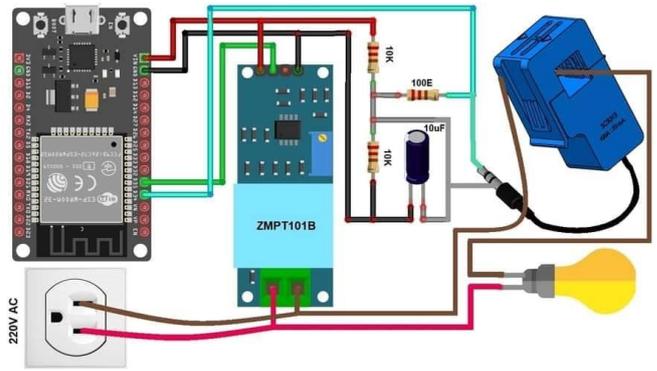

Fig. 2. Circuit diagram

### B. Algorithm

- Readings (voltage, current) extracted from connected appliance using voltage and current sensors
- Processed in ESP32 WROOM module for cost calculation
- Sent to BLYNK mobile app (third-party) for display.
- BLYNK app allows remote on/off control via relay module
- Readings stored in Firebase runtime server
- Data populates React webpage automatically
- Changes in Firebase reflect instantly in React webpage
- Monthly submission of webpage
- Metamask interface for transaction info and gas
- Initiate payment in Metamask, transaction begins
- After a few seconds, transaction completes
- View blockchain block addition in Ganache software
- Data secured and immutable with blockchain

### C. Formula

The "emon" object is used to measure the perceived power of the electricity usage and is used to create the formula for estimating electricity consumption in kilowatt-hours (kWh). The equation is

$$kWh = kWh + (emon.apparentPower * (millis() - lastmillis)) / 3600000000.0 \quad (1)$$

The "emon.apparentPower" in the formula stands for the apparent power in Watts (W), which is made up of both active (real) power and reactive power. The next section, (millis() - lastmillis), determines how many milliseconds have passed since the last measurement was logged. This period denotes the period of time that the apparent power was measured.

The formula uses the constant value 3600000000.0, which stands for the amount of milliseconds in an hour (60 seconds * 60 minutes * 1000 milliseconds), to convert the time difference from milliseconds to hours. Hours are derived by multiplying the time difference by this constant.

The updated value of kWh is then created by multiplying the current cumulative energy consumption (kWh) by the apparent power multiplied by the time difference (in hours). This makes it possible to track and update the kWh value continuously when fresh power readings are acquired. The formula offers a useful and effective way to gauge and monitor power use, making it easier to control energy costs and monitor usage in real-time.

*D. Hardware*

TABLE 2   HARDWARES USED

| Sr.no | Hardware | Work |
|---|---|---|
| 1. | ESP WROOM 32 | Main component which will help to process the information and pass on to the software. |
| 2. | Voltage Sensor (ZMPT101B) | To collect the present voltage. |
| 3. | Current Sensor (SCT-013) | To collect the present current flowing through the wire. |
| 4. | Resistors (10k and 100E) | To help the flow of current in circuit. |
| 5. | Capacitor | To maintain the voltage level. |
| 6. | Relay Module | To turn on/off the appliance connected. |
| 7. | Breadboard | To place all the components. |
| 8. | Jumper wires | To connect components with each other |

The IoT-based Electricity Energy Monitor utilising ESP32 cannot be implemented without the components provided. Data is gathered and processed centrally by the primary component, ESP WROOM 32, before being sent to the software interface. A vital component in gathering the current voltage is the Voltage Sensor ZMPT101B, while the Current Sensor SCT-013 records the current flowing through the wire in real time. A 10k resistor and a 100E resistor, among other resistors, help to control the current flow inside the circuit. A capacitor is also used to maintain and stabilise the voltage levels. The Relay Module enables the turning on and off of the connected appliance to regulate its functioning.

The Breadboard provides a physical architecture for the system and acts as a foundation for connecting and organising all the parts. Jumper wires are used to provide connectivity between the various components. This collection of parts works together to provide an effective and useful IoT-based electricity energy monitor that can detect voltage and current accurately and regulate appliances.

*E. Softwares*

TABLE 3   SOFTWARES USED

| Sr no | Software | Version | Work |
|---|---|---|---|
| 1. | Arduino IDE | 1.8.19 | To code the esp-32 processor. |
| 2. | Blynk application | 2.27.32 | GUI for user to see the live consumption. |
| 3. | Ganache | 7.0.0 | Blockchain development tool |
| 4. | Visual Studio Code | 1.60 | IDE used to integrate everything |
| 5. | Firebase | 9.0.0 | Used as database |
| 6. | React | 17.0.2 | JavaScript library used to implement the website |

The IoT-based Electricity Energy Monitor system's implementation depends heavily on the software components provided. The ESP32 processor may be programmed using the Arduino IDE version 1.8.19, which also allows for the development of the device's logic and functionality. Users may view real-time statistics on power use using the user-friendly graphical interface provided by the Blynk programme version 2.27.32. The use of Ganache version 7.0.0 as a development tool for blockchain integration makes it easier to construct safe and open transactions for energy usage. Version 1.60 of Visual Studio Code serves as the integrated development environment (IDE) where all components are seamlessly connected, speeding up the process of system integration as a whole. The database, which is Firebase version 9.0.0, effectively stores and manages the data produced by the system. Additionally, the system benefits from React version 17.0.2, a JavaScript package that makes it possible to build interactive, dynamic websites that improve user experience. The IoT-based Electricity Energy Monitor system is driven by various software components working together to provide coding capabilities, user interaction interfaces, blockchain functionality, seamless integration, efficient data storage, and dynamic web display.

*F. Working*

On a breadboard, several parts of the hardware configuration are linked to one another. Notably, a light serving as a metaphor for an appliance is connected to the current and voltage sensors. The light is turned on, and the readings that appear are carefully examined. These readings offer information about the linked appliance's power use. These results are shown on the Blynk interface, which also provides real-time statistics on the amount of power used by any other circuit-integrated devices as well as the lightbulb. In this configuration, realtime power usage data is visualised using the Blynk platform, increasing user awareness and enabling effective energy management. This setup provides a real-world example of energy monitoring.

The current and voltage sensors, which are linked to a bulb, are the only components of the IoT-based Electricity Energy Monitor (SEM) that are not integrated into a breadboard during hardware setup. Readings may be directly analysed with this configuration. Real-time measurements are taken when the light is turned on and stored for further examination. Visualising the power consumption measurements of the lightbulb or any other appliance connected to the circuit is made possible by the Blynk interface. This user-friendly interface acts as a dashboard that shows the current energy usage, promoting a thorough awareness of energy consumption trends. Additionally, users may easily turn the bulb on or off from a distance using the Blynk application.

Integrating Firebase, a real-time database, is essential for handling and storing the dynamic data produced by the electrical components. The readings from the numerous sensors are logged and organised in this database, guaranteeing reliable and easy-to-access data retrieval. A specialised website serves as a portal for data integration into blockchain technology in addition to the SEM. Through the use of the software application Metamask, SEM data may be seamlessly added to the blockchain, improving security and transparency through decentralised record-keeping.

The Ganache interface, which displays the addition of blocks to the blockchain, is a crucial part of the SEM ecosystem. It displays the timestamp, the amount of gas used for each transaction, and the hash code related to the block contents. Due to the traceability and verifiability of every block added to the blockchain, this openness provides responsibility in the data recording process. The integrity and dependability of the data saved in the blockchain are further reinforced by Ganache, which draws attention to the values of the recorded data within each block.

To sum up, the IoT-based SEM configuration integrates hardware and software components in a fluid manner, providing real-time monitoring of energy use using the Blynk interface. Data management and security are improved by the combination of Firebase and blockchain technology, and the Ganache user interface shows actual blockchain activity. This complete solution gives customers the capability to remotely manage appliances and take part in the safe, decentralised collection of energy consumption data in addition to providing effective energy monitoring.

## IV. ELECTRICITY CONSUMPTION FORECASTING

### A. Prophet model

Facebook's Core Data Science team created the Prophet model, a simple time series forecasting method. It excels in streamlining forecasting operations while preserving precision and versatility. It offers automated identification of numerous seasonal patterns such daily, weekly, and yearly cycles by breaking down time series data into trend, seasonality, and holiday components. Notably, the model takes vacations and other special occasions into account, enabling users to enter known impacts and modify projections appropriately. Its adaptability includes both logistic and linear trend modelling, making it suitable for a range of data patterns. Prophet also manages missing data and outliers, ensuring resilience in noisy datasets. The clear methodology of the concept offers insights into each contributing aspect, facilitating interpretation. The Prophet model is an invaluable tool for forecasting since it is scalable and applicable to numerous time series.

In its simplest version, the Prophet model is expressed as follows:

$$y(t) = g(t) + s(t) + h(t) + X(t) + \varepsilon \qquad (2)$$

The observed value of the time series at time t is represented by y(t).
The trend component is g(t).
The seasonality factor is denoted by s(t).
The holiday effect component is h(t).
The extra regressors (if any) are represented by X(t).
The error term is represented by $\varepsilon$.

### B. Data manipulation & transformation

For the Prophet model to produce reliable predictions in the area of forecasting home electricity use, the manipulation and transformation of raw data are crucial steps. To preserve temporal coherence while estimating missing values, a number of imputation strategies are investigated, starting with rigorous missing value handling. These techniques include forward-fill, backward-fill, and interpolation. Consequently, to find and eliminate probable outliers that can skew patterns, reliable outlier detection techniques are used, such as interquartile range and Z-score. Inconsistencies, duplication, and data integrity are addressed by a thorough data cleaning procedure, and consistent time intervals are achieved via temporal resampling, allowing the collection of fine-grained temporal consumption patterns. Moving averages and STL, for example, untangle the data into intrinsic components, showing long-term trends and seasonal patterns crucial for the Prophet model. Seasonality and trend decomposition techniques, such as STL and moving averages, do the same. To improve forecast accuracy, feature engineering uses domain expertise to incorporate external factors such as weather data and demographic information. Training and testing data are kept in the same temporal sequence thanks to careful data splitting and normalization procedures that standardize feature scales. These several phases work together to create a polished dataset that is ready to provide accurate, noise-resistant forecasting in the context of residential power usage, eventually supporting energy management strategies and resource allocation choices.

### C. Model configuration & parameter selection

Careful parameter setup is necessary for reliable forecasts when using the Prophet model to simulate residential power usage. This technique necessitates a nuanced comprehension of fundamental elements.
Seasonality and Trend: It is essential to have a thorough understanding of seasonality (daily, weekly, and annual cycles) and trend (linear, logistic). Consumption habits and the trajectory of growth are influenced by these factors.
Holiday Effects: Given their large influence on consumption, holidays must be taken into account in the model. Custom holiday dataframes record particular behaviour, and holidays_prior_scale controls how much of an impact they have.

Additional Regressors: Accuracy can be improved by include external factors such as weather, economy, and events. The addition of these using add_regressor makes it easier to recognise intricate patterns.

Parameter choices: Parameter tweaking is crucial. The relationship between seasonality and trend is determined by seasonality_mode. Adjusting sensitivity to trend changes is done with changepoint_prior_scale. Holiday impact intensity is controlled by holidays_prior_scale.

By allowing the Prophet model to extract and anticipate specific details of use, this complex arrangement maximises its potential and improves energy management and well-informed decision-making.

### D. Benefits of prophet model

Compared to other time series forecasting methods, the Prophet model has strong benefits. Preprocessing duties are made easier by its user-friendly architecture, ability to manage missing data and outliers, and seamless integration of holidays and events. The model is adaptable for many data circumstances due to its automated recognition of numerous seasonality and flexibility in trend identification. Due to its openness, understanding is aided by insights into forecasting components. Real-time forecasting capabilities are appropriate for dynamic applications, while the short prototyping methodology encourages experimentation. Prophet's adaptability to various time series, open-source status, and vibrant community support further guarantee its applicability and availability. Prophet's mix of characteristics, which is especially helpful for non-experts, making it a potent tool for precise and understandable time series forecasting across diverse fields.

### E. Application and future direction

Through the application of sophisticated machine learning algorithms to decipher complex consumption patterns, this work highlights the potential to advance time series forecasting for residential power usage. Additionally, it promotes the merging of outside data sources to improve forecast accuracy. The research builds a realistic framework for data modification, transformation, and analysis using historical Firebase data, clearly illustrating the Prophet model's power to make precise predictions. Its dependability is strengthened by its clear methodology, skillful management of missing data, inclusion of holiday impacts, and user-friendly implementation. The study specifically acknowledges the limitation of restricted data availability, which led to the selection of initial parameters, with future intentions to further develop the system. This strategy has a lot of potential for cost reduction, energy management, and ecological sustainability. The research emphasizes the practical value of precise energy consumption forecasting by outlining a thorough approach and demonstrating its virtues, which encourages better resource allocation and well-informed decision-making in homes and utilities.

### V. CONCLUSION

The adoption of blockchain technology in the power billing system presents a game-changing opportunity to tackle several issues and usher in a new age of efficiency and transparency. We provide an immutable ledger where metre readings are safely kept and resistant to manipulation or unauthorized changes by easily integrating blockchain. By doing away with the laborious procedure of taking manual metre readings, this innovative method reduces the likelihood of mistakes and inaccuracies caused by humans that frequently occur in conventional systems.

The use of smart contracts to this blockchain architecture also advances the billing process into an automated and effective space. Smart contracts automate billing computations and transactions using pre-programmed criteria and actions, removing the need for manual involvement and its accompanying delays. This simplified strategy ensures that customers are invoiced fairly and properly while also improving operational efficiency.

The benefits go beyond simple billing effectiveness. Giving consumers the means to keep track of their daily energy usage provides priceless data that can encourage more responsible energy usage behaviors. A more sustainable environment and less energy waste may result from decisions regarding energy usage that are guided by this increased understanding.

Additionally, the incorporation of blockchain technology lowers costs all around. Operational costs are greatly decreased by doing away with labor-intensive processes like manual reading and processing. This results in financial gains for electricity providers and customers alike.

In conclusion, the adoption of blockchain technology has the potential to completely alter the way that power is billed. The system becomes more secure, effective, and user-centric by creating an unchangeable record of metre readings and automating billing procedures using smart contracts. Giving customers access to real-time data encourages prudent energy use, fostering a cleaner future. The combination of efficiency, security, and transparency presents blockchain as a catalyst for improvement in the power billing industry, providing significant advantages to all parties involved.